\documentclass[twocolumn,pra,aps,amssymb]{revtex4}
\usepackage{algorithm2e}
\usepackage{amsmath,amssymb,amsfonts,amsthm}
\usepackage{multirow}
\usepackage{verbatim}
\usepackage{url}
\usepackage{comment}
\usepackage{mathtools}
\usepackage{epsf,pgf,graphicx}
\usepackage{color}
\usepackage{tikz,pgf}

\begin{document}
\newcommand{\ket}[1]{\ensuremath{\left|#1\right\rangle}}
\newcommand{\bra}[1]{\ensuremath{\left\langle#1\right|}}
\newcommand\floor[1]{\lfloor#1\rfloor}
\newcommand\ceil[1]{\lceil#1\rceil}
\newtheorem{definition}{Definition} 
\newtheorem{theorem}{Theorem} 
\newtheorem{claim}{Claim}
\newtheorem{lemma}{Lemma}
\title{Device Independent Quantum Private Query}
\author{Arpita Maitra$^1$, Goutam Paul$^2$ and Sarbani Roy$^3$}
\affiliation{$^1$Indian Institute of Management Calcutta, India.\\
Email: arpita76b@gmail.com\\
$^2$Indian Statistical Institute, Kolkata.\\
Email: goutam.paul@isical.ac.in\\
$^3$Indian Institute of Technology Kharagpur, India.\\
Email: sarbani16roy@gmail.com\\
}

\begin{abstract}
In Quantum Private Query (QPQ), a client obtains values corresponding to his query only and nothing else from the server and the server does not get any information about the queries. Giovannetti et al. (Phys. Rev. Lett., 2008) gave
the first QPQ protocol and since then quite a few variants and extensions have been proposed. However, none of the existing protocols are device independent, i.e., all of them assume implicitly that the entangled states supplied to the client and the server are of certain form. In this work, we exploit the idea of a local CHSH game and connect it with the scheme of Yang et al. (Quantum Inf. Process., 2014)  to present  the concept of device independent QPQ protocol for the first time.
\end{abstract}
\maketitle

\section{Introduction}
During the last two decades, Quantum Key Distribution (QKD) has remained the main theme of quantum cryptography. In recent times, however, several other quantum cryptographic primitives are being explored and 
Quantum Private Query (QPQ) is one of them. In QPQ, a client issues queries to a database and obtains the real values without knowing anything else about the database, whereas the server should not gain any information about the queries. Here, we assume that Bob is the database holder or server and Alice is the client.
The first protocol in this domain had been proposed by  Giovannetti et al.~\cite{GLM} followed by ~\cite{GLM11} and~\cite{Olejnik}. However, those scheme are highly theoretical and difficult for implementation. For implementation purpose, Jakobi et al.~\cite{jakobi} came out with a QPQ protocol which was based on SARG04 quantum key distribution protocol~\cite{SARG}. In 2012, Gao et al.~\cite{Gao} proposed a flexible generalization of~\cite{jakobi}. Rao et al.~\cite{Rao} suggested two more efficient modifications of classical post-processing in the protocol of Jakobi et al. In 2013, Zhang et al.~\cite{Zhang} proposed a QPQ protocol based on counterfactual QKD scheme~\cite{Noh}. In 2014, Yang et al. came out with a flexible QPQ protocol~\cite{Yang} which was based on B92 quantum key distribution scheme~\cite{b92}. This domain is gradually improving. It is evident from the large number of published literatures~\cite{Wei,Chan,Gao2,Liu} in the recent two years.

The security of all those protocols is defined on the basis of the following facts.

(a) Bob knows the whole key which would be used for the encryption of the database.

(b) Alice knows a fraction of bits of the key.

(c) Bob does not get any information about the position of the bits which are known to Alice.

Thus, it is vary natural that in QPQ protocol, there is no need for an outsider adversary. Unlike QKD, here, one of the legitimate parties is playing the role of an adversary. Alice tries to extract more information about the raw key bits, whereas Bob tries to know the position of the bits known to Alice. 

We identify that the security of all the existing protocols are based on the fact that Bob relies his devices, i.e., the source which supplies the qubits and the detectors which measure the qubits. Thus, similar to the QKD protocols, trustworthiness of the devices are implicit in the security proofs of the protocols. In the current work, we try to understand if we remove such trustworthiness from the devices like Device Independent QKD~\cite{yao,acin06a,acin06b,scarani06,Acin}.

In DI-QKD, a statistical test known as Bell test~\cite{Bell} or CHSH test~\cite{CHSH} is performed to verify whether the shared entangled states between the legitimate parties are maximally entangled. If the states are maximally entangled, then QKD protocol provides unconditional security. However, the test has to be performed non-locally. In other words, two distant parties (Alice and Bob) have to be involved in CHSH test.

Very recently, Lim et al.~\cite{Lim} proposed a DI-QKD scheme where they exploit the idea of local CHSH test. In local CHSH test, the sender performs CHSH test at his or her end in the motivation towards certifying whether the states, going to be used for QKD, are maximally entangled. 

In case of QPQ, we identify that if the states shared between Bob and Alice are not in a certain form, then Alice can always apply some strategies which help her to extract more information about the raw key bits than what is suggested by the protocol. Thus, it is necessary for Bob to certify whether the states are in the desired form. Motivated by the idea of local CHSH test by Lim et al.~\cite{Lim}, we, here, propose a protocol which provides this certification. 
The value obtained from the test will depend upon the predefined success probability of Alice about the raw key bits. In other words, how much information about the key has to be allowed to Alice by the protocol.

Here, we work on the QPQ protocol presented by Yang et al.~\cite{Yang}.   
Note that this protocol~\cite{Yang} can be performed by a certain kind of mixed state also (namely, $\left(\ket{0,\phi_0}\bra{0,\phi_0}+\ket{1,\phi_1}\bra{1,\phi_1}\right)/2$). Since we try to prove the device independence security by exploiting the idea of local CHSH test, we work with the entanglement-based version. Further, our suggested scheme for testing device independence lies on top of the QPQ protocol of~\cite{Yang}. One can replace that QPQ part by any other entanglement-based QPQ protocol (which might not be performed by any mixed state) and our scheme will work there too.

QPQ protocol can be viewed as a two-party (mistrustful) cryptography, i.e., two parties that want to perform a certain task together without fully trusting
each other. In this regard, one may mention~\cite{Silman1,Silman2,Kaniewski1,Kaniewski2,Wehner} where similar ideas have been exploited to propose imperfect coin flipping and bit commitment in DI setting.

We first revisit the protocol of Yang et al.~\cite{Yang}. Next, we show how Alice could choose a strategy to extract more information about the raw key bits if the shared entanglement between her and Bob is not in a desired form. We, then, come out with the idea of local CHSH like test which is exploited to certify whether the states are secure for QPQ protocol. All the lemmas and theorems used to prove the security of the proposed protocol are given in Appendix A. \medskip

\section{Revisiting the Protocol of~\cite{Yang}}
In this section we revisit the protocol for quantum private query proposed in~\cite{Yang}.  The protocol exploits the idea of B92 quantum key distribution scheme. There are two phases in the protocol, namely, key generation and private query. In the key generation phase,
Bob and Alice share entangled states of the form
$\frac{1}{\sqrt{2}}(\ket{0}_{B}\ket{\phi_0}_{A}+\ket{1}_{B}\ket{\phi_1}_{A})$, where, $\ket{\phi_{0}}_{A}=\cos{(\frac{\theta}{2})}\ket{0}+\sin{(\frac{\theta}{2})}\ket{1}$ and $\ket{\phi_{1}}_{A}=\cos{(\frac{\theta}{2})\ket{0}}-\sin{(\frac{\theta}{2})}\ket{1}$. Here, subscript B stands for Bob and subscript A stands for Alice. $\theta$ may vary from $0$ to $\frac{\pi}{2}$.
After receiving the qubits from Bob, Alice announces the position of the qubits that have ultimately  reached at the end of Alice. Bob discards the lost photons. 
After post selection, Bob measures his qubits in $\{\ket{0}_{B}, \ket{1}_{B}\}$ basis, whereas Alice measures her qubits either in $\{\ket{\phi_{0}}_{A},\ket{\phi_{0}^{\perp}}_{A}\}$ basis or in $\{\ket{\phi_{1}}_{A}, \ket{\phi_{1}^{\perp}}_{A}\}$ basis randomly.  
If the measurement result of Alice gives $\ket{\phi_{0}^{\perp}}$, she concludes that the raw key bit at Bob's end must be $1$. If it would be $\ket{\phi_{1}^{\perp}}$, the raw key bit must be $0$. 
Bob and Alice execute classical post-processing so that Alice's information on the key reduces  to one bit or more.  Bob knows the whole key, whereas Alice generally knows several bits of the key.   

In the private query phase,
if Alice knows the $j$th bit of the key $K$ and wants to know the $i$th element of the database, she declares the integer $s=j-i$. 
Bob shifts $K$ by $s$ and hence gets a new key, say $K_0$.
Bob encrypts his database by this new key $K_0$ with one-time pad and sends the encrypted database to Alice.
 Alice decrypts the value with her $j$th key bit and gets the required element of the database.

The security of the protocol comes from the fact that Alice knows the final key partially. Thus, even if she gets  access to the whole encrypted database, she can not obtain the full information about the database. 
Now, we will calculate the success probability of Alice to guess a bit in raw key.

As Bob measures his qubits only in $\{\ket{0}_{B}, \ket{1}_{B}\}$ basis, he will get either $\ket{0}$ with probability $\frac{1}{2}$
or $\ket{1}$ with probability $\frac{1}{2}$. When Bob gets $\ket{0}$, Alice should get $\ket{\phi_0}$. If she chooses $\{\ket{\phi_0}_{A}, \ket {\phi_0^{\perp}}_{A}\}$ basis, she will get $\ket{\phi_0}$ with probability $1$ and never gets $\ket{\phi_0^{\perp}}$. However, if she chooses $\{\ket{\phi_1}_{A}, \ket{\phi_1^{\perp}}_{A}\}$ basis, she will get either $\ket{\phi_1}$ with probability $\cos^2\theta$ or $\ket{\phi_1^{\perp}}$ with probability $\sin^2\theta$. We formalize all the conditional probabilities in the following table.

{\scriptsize
\begin{center}
\begin{tabular}{|c|c|c|c|c|}
\hline
 & \multicolumn{4}{|c|}{Cond. Probability of Alice}\\
\cline{2-5}
& A=$\ket{\phi_0}$ & A=$\ket{\phi_0^{\perp}}$ & A=$\ket{\phi_1}$ & A=$\ket{\phi_1^{\perp}}$\\
\hline
$ B=0 $ & $\frac{1}{2}.1$&$\frac{1}{2}.0$ &$\bf{\frac{1}{2}.\cos^2\theta}$ & $\bf{\frac{1}{2}.\sin^2\theta}$\\
\hline
$B=1$ & $\bf{\frac{1}{2}.\cos^2\theta}$ & $\bf{\frac{1}{2}.\sin^2\theta}$ &$\frac{1}{2}.1$ &$\frac{1}{2}.0$\\
\hline
\end{tabular}
\end{center}}

According to the protocol, when Alice gets $\ket{\phi_0^{\perp}}$, she outputs $1$. And when she gets $\ket{\phi_1^{\perp}}$, she outputs $0$. Thus, the success probability of Alice to guess a bit in raw key can be written as 
{\scriptsize
$\Pr(A=B)$
\begin{eqnarray}
\label{sucprob}
&=&\Pr(A=0,B=0)+\Pr(A=1,B=1) \nonumber\\
&=&\Pr(B=0).\Pr(A=0|B=0)+\Pr(B=1).\Pr(A=1|B=1)\\
&=&\frac{1}{2}.\Pr(A=\phi_1^{\perp}|B=0)+\frac{1}{2}.\Pr(A=\phi_0^{\perp}|B=1). \nonumber
\end{eqnarray}}
From the above table, we can see that the success probability of Alice becomes $\frac{\sin^2\theta}{2}$.

\section{Biased Choice of Alice's Basis}
Suppose, Bob trusts the source, i.e., he believes that the states shared between Alice and him are of the certain form~\cite{Yang}. Let, the source supplies some arbitrary entangled states $(\alpha\ket{0}_{B}\ket{\phi_0}_{A}+\beta\ket{1}_{B}\ket{\phi_1}_{A})$, where $|\alpha|^2=(\frac{1}{2}+\epsilon)$ and
$|\beta|^2=(\frac{1}{2}-\epsilon)$ to Bob. Suppose, Alice has this information and also the information about the values of $\alpha$ and $\beta$. In this case, she chooses the basis as follows.
\begin{itemize}
\item $\{\ket{\phi_0}_{A}, \ket{\phi_0^{\perp}}_{A}\}$ with probability $\frac{1}{2}-\epsilon$.

\item $\{\ket{\phi_1}_{A}, \ket{\phi_1^{\perp}}_{A}\}$ with probability $\frac{1}{2}+\epsilon$.
\end{itemize}
 
Her success probability can be calculated from the following table.

{\tiny
\begin{center}
\begin{tabular}{|c|c|c|c|c|}
\hline
 & \multicolumn{4}{|c|}{Cond. Probability of Alice}\\
\cline{2-5}
& A=$\ket{\phi_0}$ & A=$\ket{\phi_0^{\perp}}$ & A=$\ket{\phi_1}$ & A=$\ket{\phi_1^{\perp}}$\\
\hline
$ B=0 $ &$ 1.(\frac{1}{2}-\epsilon)$& 0 &$(\cos^2\theta)\cdot\left(\frac{1}{2}+\epsilon\right)$ & $(\sin^2\theta)\cdot\left(\frac{1}{2}+\epsilon\right)$\\
\hline
$B=1$ & $(\cos^2\theta)\cdot\left(\frac{1}{2}-\epsilon\right)$ & $(\sin^2\theta)\cdot\left(\frac{1}{2}-\epsilon\right)$ &$1\cdot\left(\frac{1}{2}+\epsilon\right)$ & 0\\
\hline
\end{tabular}
\end{center}
}
Following Eq.~\eqref{sucprob}, it becomes
$(\frac{1}{2}+2\epsilon^2)\sin^2\theta$. 

Thus, if Alice and Bob do not share the entangled states of the certain kind, then Alice can always extract more information about the raw key bit following the suggested strategy. The biasing on the bases of Alice depends on the values of $\alpha$ and $\beta$. For example, if $\alpha=\frac{1}{2}-\epsilon$ and $\beta=\frac{1}{2}+\epsilon$, then Alice chooses $\{\ket{\phi_0}_{A}, \ket{\phi_0^{\perp}}_{A}\}$ with probability $\frac{1}{2}+\epsilon$ and chooses $\{\ket{\phi_1}_{A}, \ket{\phi_1^{\perp}}_{A}\}$ with probability $\frac{1}{2}-\epsilon$.

To mitigate such problem, Bob has to remove his trust from the devices and has to perform some local test at his end to become sure that the states shared between them are of the specific form~\cite{Yang}. As we consider the entanglement version of the QPQ protocol, we suggest a local statistical test which is actually CHSH test performed locally. The difference is that for this test, we do not require the perfect CHSH value. The value depends on the value of $\theta$.

However, when the states are of the form given in~\cite{Yang}, then the above strategy does not help Alice to extract more information about the raw key bit. Let Bob and Alice share the entangled states of the specific form and Alice chooses her measurement bases $\{\ket{\phi_0}_{A},\ket{\phi_0^{\perp}}_{A}\}$ and $\{\ket{\phi_1}_{A},\ket{\phi_1^{\perp}}_{A}\}$ with probability $\frac{1}{2}-\epsilon$ and $\frac{1}{2}+\epsilon$ respectively. In this case, following Eq.~\eqref{sucprob}, the success probability becomes
$\frac{\sin^2\theta}{2}$.

Thus, it will be necessary for Bob to certify that those shared states are of the certain form. In the following section we propose a protocol which certify this. Thus, Bob no longer requires to put trust on the source as well as the detectors. By performing a test which is almost like CHSH test at his end, he first checks whether the states follow the desired property. Conditioning on the success of the test, Bob proceeds for QPQ. Here, we consider detectors with unit efficiency. However, for practical implementation of the suggested protocol, one has to consider the detectors with non-unit efficiency. 

One may wonder why we have not chosen quantum state tomography to check whether the states are of the certain form. The reason is that tomography would require an infinite number of states to achieve perfect accuracy. On the other hand, choosing a different avenue of local CHSH game, we are able to analyse the security of our protocol for finite number of states.

\section{Our Protocol and Local CHSH Game}
Before describing the proposed protocol, we first enumerate the assumptions required for the security of the protocol. Those are summarized as follows.

1. Devices are causally independent, i.e., each use of the device is independent of the previous use. This assumption implies that the devices are memoryless.

2. Alice and Bob's laboratories are perfectly secured i.e., no information is leaked from their laboratories. 

3. All the detectors at Bob's end have unit efficiency i.e., he always gets conclusive outcomes. 

Our protocol is described in Algorithm~1. For brevity, we write $\gamma n$ and $(1-\gamma) n$ instead of $\lceil{\gamma n}\rceil$ and $\lfloor{(1-\gamma) n}\rfloor$ respectively.

\restylealgo{boxed}
\begin{algorithm}[htbp]
{\scriptsize
\begin{enumerate}
\item Bob starts with $n$ number of entangled states.
\item Bob divides the given entangled pairs into two sets. One is $\Gamma_{CHSH}$ and another is $\Gamma_{QPQ}$. The set $\Gamma_{CHSH}$ contains $\gamma n$ number of entangled states, whereas $\Gamma_{QPQ}$ contains $(1-\gamma) n$ number of the entangled states for $0<\gamma<1 $. 
 \item For rounds $i \in \{1,\cdots, \gamma n\}$

\hspace{5pt} (a) Bob chooses $x_i\in\{0,1\}$ and $y_i\in\{0,1\}$ uniformly at random. 

\hspace{5pt} (b) If $x_i=0$, he measures the first particle of the entangled state in $\{\ket{0}, \ket{1}\}$ basis and if $x_i=1$, he measures that in $\{\ket{+}, \ket{-}\}$ basis. 

\hspace{5pt} (c) Similarly, if $y_i=0$, Bob measures the second particle of the entangled state in $\{\ket{\psi_1}, \ket{\psi_1^{\perp}}\}$ basis and if $y_i=1$, he measures that in $\{\ket{\psi_2}, \ket{\psi_2^{\perp}}\}$ basis. 

\hspace{5pt} (d) The output is recorded as $a_i (b_i) \in\{0,1\}$ for the first (second) particle. 
The encoding for $a_i (b_i)$ is as follows. 
\begin{itemize}
\item For the first particle of each pair, $a_i=0$ if the measurement result is $\ket{0}$ or $\ket{+}$; it is 1 if the result would be $\ket{1}$ or $\ket{-}$.
\item For the second particle of each pair, $b_i=0$ if the measurement result is $\ket{\psi_1}$ or $\ket{\psi_2}$; it is 1 if the measurement result would be $\ket{\psi_1^{\perp}}$ or $\ket{\psi_2^{\perp}}$, then $b_i=1$.
\end{itemize}

\hspace{5pt} (e) Testing: For the test round $i \in \Gamma_{CHSH}$, define 
  \begin{eqnarray*}
  Y_i=
  \begin{cases}
   1 & \text{if } a_i\oplus b_i=x_i \wedge {y_i}\\
   0 & \text{if } otherwise.
  \end{cases}
 \end{eqnarray*} 
 \item If 
 $\frac{1}{\gamma n}\sum_i {Y_i} 
 < \frac{1}{8}(\sin{\theta}(\sin\psi_1+\sin\psi_2)+\cos\psi_1-\cos\psi_2)+\frac{1}{2}$,
  Bob aborts the protocol. 
  \item Conditioning on the event that the local CHSH test at Bob's end has been successful, Bob proceeds for the subset $\Gamma_{QPQ}$ and sends one halves of the remaining $(1-\gamma) n$ number of entangled pairs to Alice.  
 \item Alice performs the private query phase as in~\cite{Yang}.
 \end{enumerate}
}
\caption{Our Proposed protocol, $\bf \Pi$}
\end{algorithm}

\begin{figure*}[htbp]
\centering
\begin{tabular}{cc}
\includegraphics[width=0.45\textwidth]{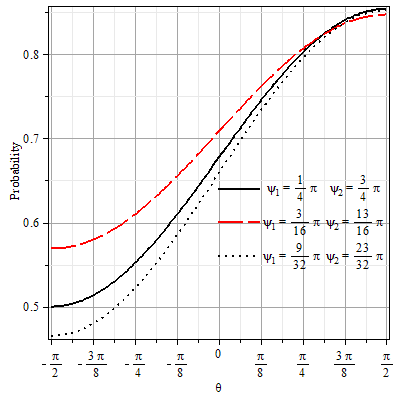}
& 
\includegraphics[width=0.45\textwidth]{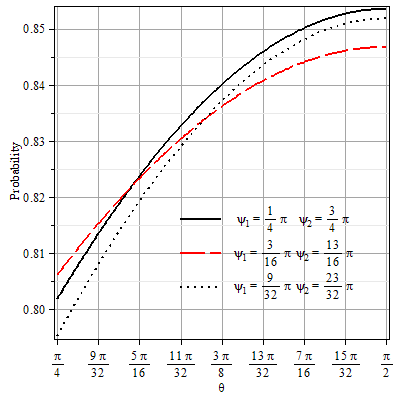}
\end{tabular}
\caption{The value of $\Pr(a_i\oplus b_i=x_i\wedge y_i)$ with respect to $\theta$}
\label{plot}
\end{figure*}

Note that we are dealing with several bases, namely, $\{\phi_0,\phi_0^{\perp}\}$, $\{\phi_1,\phi_1^{\perp}\}$, $\{\psi_1,\psi_1^{\perp}\}$ and $\{\psi_2,\psi_2^{\perp}\}$. It should be clarified that where $\{\phi_0,\phi_0^{\perp}\}$, $\{\phi_1,\phi_1^{\perp}\}$ bases are chosen by Alice for QPQ protocol, $\{\psi_1,\psi_1^{\perp}\}$ and $\{\psi_2,\psi_2^{\perp}\}$ bases are chosen by Bob to perform local CHSH test. Here, we consider {\bf{$\ket{\psi_1}=\cos{\frac{\psi_1}{2}}\ket{0}+\sin{\frac{\psi_1}{2}}\ket{1}$}} and {\bf {$\ket{\psi_2}=\cos{\frac{\psi_2}{2}}\ket{0}+\sin{\frac{\psi_2}{2}}\ket{1}$}}.

In the QPQ protocol of~\cite{Yang}, Bob measures his particles in $\{\ket{0},\ket{1}\}$ basis only. Hence, the protocol can be performed by the mixed states also. One may think that for the local CHSH game, here, it is sufficient to measure the first particle of Bob in $\{\ket{0},\ket{1}\}$ basis only as Bob does not need to test coherence (purity) of the states. However, note that our proposal for local CHSH game lies on the top of the QPQ protocol presented by Yang et al~\cite{Yang}. One can replace the QPQ part by any other entanglement based QPQ protocol which might not be performed by mixed states. Hence, it is necessary to use $\{\ket{+},\ket{-}\}$ basis in proposed CHSH test.    

Next, we analyze case by case situation of the proposed CHSH like test. Let Bob obtains the entangled states of the form $\frac{1}{\sqrt{2}}(\ket{0}_{B}\ket{\phi_0}_{A}+\ket{1}_{B}\ket{\phi_1}_{A})$. We calculate the conditional probabilities for each case and present those in a tabular form in Table~\ref{tab1}.

Since $\Pr(x_i,y_i) = \frac{1}{4}$ for all $x_i, y_i$, 
multiplying each individual probability in Table~\ref{tab1} by
$\frac{1}{4}$ gives the corresponding joint probabilities.
We have 
{\scriptsize
\begin{eqnarray*}
\Pr(a_i \oplus b_i &=& x_i \wedge y_i) =\\
  \Pr((x_i,y_i) &=& (0,0) \ \& \ ((a_i,b_i) = (0,0) \ OR \ (1,1)))\\
 + \Pr((x_i,y_i) &=& (0,1) \ \& \ ((a_i,b_i) = (0,0) \ OR \ (1,1))) \\
+ \Pr((x_i,y_i) &=& (1,0) \ \& \ ((a_i,b_i) = (0,0) \ OR \ (1,1)))\\
+ \Pr((x_i,y_i) &=& (1,1) \ \& \ ((a_i,b_i) = (0,1) \ OR \ (1,0))).
\end{eqnarray*}}
Adding the joint probabilities for the corresponding rows,
we find that the above quantity is equal to $\frac{1}{8}[\sin \theta (\sin \psi_{1} +\sin \psi_{2} )+(\cos \psi_{1}-\cos \psi_{2})]+\frac{1}{2}$.

\begin{table}[htbp]
\caption{Conditional probability of $(a_i, b_i)$ given $(x_i, y_i)$}
\centering
\begin{tabular}{|c|c|c|}
\hline
$(x_i, y_i)$ & $(a_i, b_i)$ &$\Pr\left((a_i,b_i) \ | \ (x_i, y_i)\right)$\\
\hline
\multirow{4}{*}{(0, 0)} & (0, 0) & $\frac{1}{2} \cos^{2}(\frac{\theta-\psi_{1}}{2})$ \\
\cline{2-3}
			& (0, 1) &  $\frac{1}{2} \sin^{2}(\frac{\theta-\psi_{1}}{2})$\\
\cline{2-3}
			& (1, 0) & $\frac{1}{2} \cos^{2}(\frac{\theta+\psi_{1}}{2})$  \\
\cline{2-3}
			& (1, 1) & $\frac{1}{2} \sin^{2}(\frac{\theta+\psi_{1}}{2})$  \\
\hline
\multirow{4}{*}{(0, 1)} & (0, 0) & $\frac{1}{2} \cos^{2}(\frac{\theta-\psi_{2}}{2})$\\
\cline{2-3}
			& (0, 1) &  $\frac{1}{2} \sin^{2}(\frac{\theta-\psi_{2}}{2})$ \\
\cline{2-3}
			& (1, 0) &  $\frac{1}{2} \cos^{2}(\frac{\theta+\psi_{2}}{2})$\\
\cline{2-3}
			& (1, 1) & $\frac{1}{2} \sin^{2}(\frac{\theta+\psi_{2}}{2})$ \\
\hline
\multirow{4}{*}{(1, 0)} & (0, 0) & $\cos^{2}(\frac{\theta}{2})\cos^{2}\frac{\psi_{1}}{2}$\\
\cline{2-3}
			& (0, 1) & $\cos^{2}(\frac{\theta}{2})\sin^{2}\frac{\psi_{1}}{2}$ \\
\cline{2-3}
			& (1, 0) &  $\sin^{2}(\frac{\theta}{2})\sin^{2}\frac{\psi_{1}}{2}$  \\
\cline{2-3}
			& (1, 1) & $\sin^{2}(\frac{\theta}{2})\cos^{2}\frac{\psi_{1}}{2}$\\
\hline
\multirow{4}{*}{(1, 1)} & (0, 0) & $\cos^{2}(\frac{\theta}{2})\cos^{2}\frac{\psi_{2}}{2}$\\
\cline{2-3}
			& (0, 1) & $\cos^{2}(\frac{\theta}{2})\sin^{2}\frac{\psi_{2}}{2}$\\
\cline{2-3}
			& (1, 0) & $\sin^{2}(\frac{\theta}{2})\sin^{2}\frac{\psi_{2}}{2}$ \\
\cline{2-3}
			& (1, 1) & $\sin^{2}(\frac{\theta}{2})\cos^{2}\frac{\psi_{2}}{2}$ \\
\hline
\end{tabular}
\label{tab1}
\end{table}

In Fig.~\ref{plot}, we plot the joint probability as a function of $\theta$, for the angles $(\psi_1,\psi_2)=\{(\frac{\pi}{4},3\frac{\pi}{4}), (3\frac{\pi}{16},13\frac{\pi}{16}), (9\frac{\pi}{32},23\frac{\pi}{32})\}$. A magnified view of the plot for the region from $\theta=\frac{\pi}{4}$ to $\theta=\frac{\pi}{2}$ appears on the right part of the figure. From the plot it is observed that when $\theta=\frac{\pi}{2}$, the joint probability reaches the value equal to $\cos^2{\frac{\pi}{8}}$. 

\medskip

\section{Security Analysis} 
In this section, we prove the security of the proposed protocol. In earlier section, we showed that if the shared entangled states are not in a certain form, then Alice may extract more information than what is suggested by the protocol. So, at the beginning of the protocol either Bob has to trust devices blindly (device dependent assumption on which the security of the existing protocols depends) or he needs to test some statistical property by measuring the given entangled states (device independent assumption). The security of the proposed protocol comes from the following result.
\begin{theorem}
\label{thm1}
If for a random subset $\Gamma_{CHSH}\subset \{1,\cdots,n\}$ of size $\gamma n$, where, $\gamma>0$, the fraction of the inputs ($x_i$, $y_i$), $i\in\Gamma_{CHSH}$, which satisfy the CHSH condition i.e., $(a_i\oplus b_i=x_i\wedge y_i)$ is equal to $\frac{1}{8}(\sin{\theta}(\sin\psi_1+\sin\psi_2)+\cos\psi_1-\cos\psi_2)+\frac{1}{2}-\delta$, then for the remaining subset $\Gamma_{QPQ} \subset \{1,\cdots,n\}$ of size $(1-\gamma)n$, a fraction of inputs $(x_i,y_i)$, $i\in \Gamma_{QPQ}$, which satisfy the CHSH condition, is also equal to $\frac{1}{8}(\sin{\theta}(\sin\psi_1+\sin\psi_2)+\cos\psi_1-\cos\psi_2)+\frac{1}{2}-\delta$ with a negligible statistical deviation $\nu$. 

 Here, $\delta=\sqrt{\frac{1}{2\gamma n} \ln{\frac{1}{\epsilon_{CHSH}}}}$ and $\nu=\sqrt{\frac{(\gamma n+1)}{2\gamma^2(1-\gamma)n^2}\ln{\frac{1}{\epsilon_{QPQ}}}}$, $\epsilon_{CHSH}$ and $\epsilon_{QPQ}$ are negligibly small value.
\end{theorem} 

In the second result, we show that when $n$ is sufficiently large, then conditioned on the success of the above local CHSH test, one may proceed for QPQ protocol proposed by Yang et al.~\cite{Yang} for the remaining subset $\Gamma_{QPQ}$.
\begin{theorem}
\label{thm2}
 Conditioning on the event that local CHSH test has been successful for the subset $\Gamma_{CHSH}$, Bob can proceed for the QPQ protocol for the remaining subset $\Gamma_{QPQ}$ securely when $n\rightarrow \infty$.
 \end{theorem}

In~\cite{Yang}, the authors consider the security issues for two cases: (a) dishonest Alice and honest Bob and (b) honest Alice and dishonest Bob.
As the second phase of our protocol is the same as QPQ protocol proposed by Yang et al.~\cite{Yang}, the security issues for the second part of the current protocol remains the same. \medskip

\section{Discussion and Conclusion}
In this current draft, we propose a device independent scenario in quantum private query. Exploiting the idea of local CHSH test we show how Bob can remove his trust from devices. The proposed protocol is divided in two distinct parts. In the first part, Bob performs local CHSH test at his end. Conditioning on the event that the local CHSH test has been successful, Bob proceeds for QPQ protocol. We here worked on the QPQ protocol proposed by Yang et al.~\cite{Yang}. However, one can exploit any entanglement based QPQ protocol for the second phase of our proposed scheme. Here, we assume the detectors have unit efficiency. However, it remains open what would happen if the detectors are imperfect, i.e., having non-unit efficiency. To the best of our knowledge, the proposed protocol is the first device independent protocol in the domain of quantum private query.

\section{Appendix A: Lemmas and Proofs}
\begin{lemma}(Chernoff-Hoeffding~\cite{Chernoff})
Let $X=\frac{1}{n}\sum_i{X_i}$ be the average of $n$ independent random variables $X_1, X_2,\cdots,X_n$ with values $[0,1]$, and let $\mathbb{E}[X]=\frac{1}{n}\sum_i{\mathbb{E}[X_i]}$ be the expectation value of $X$, then for any 
$\delta>0$, we have $\Pr\left[|X-\mathbb{E}[X]| \geq \delta \right] \leq \exp(-2\delta^2 n).$
\end{lemma}
\begin{lemma}(Serfling~\cite{Serfling}) 
Let $\{x_1,x_2,\cdots,x_n\}$ be a list of values in $[a,b]$ (not necessarily distinct). Let $\overline{x}=\frac{1}{n}\sum_i x_i$ be the average of these random variables. Let $k$ be the number of random variables $X_1,X_2,\cdots,X_k$ chosen from the list without replacement. Then for any value of $\delta>0$, we have
$\Pr\left[|X-\overline{x}| \geq \delta \right] \leq \exp\left(\frac{-2\delta^2 kn}{(n-k+1)(b-a)}\right),$ where $X=\frac{1}{k}\sum_i X_i$.
\end{lemma}
\begin{lemma}(~\cite{Lim}, Corollary to Serfling Lemma)
Let $\mathbb{X}=\{x_1,x_2...x_n\}$ be a list of (not necessarily distinct) values in $[0,1]$ with the average $\mu_{\mathbb{X}}=\frac{1}{n}\sum_{i=1}x_i$. Let $\mathbb{T}$ be a subset of $\mathbb{X}$ of size $t$ with average $\mu_{\mathbb{T}}=\frac{1}{t}\sum_{i\in\mathbb{T}}x_i$. Let $\mathbb{K}$ be the remaining subset of $\mathbb{X}$ with size $k$ (i.e., $t+k=n$). If the average of the subset $\mathbb{K}$ is $\mu_{\mathbb{K}}=\frac{1}{n-t}\sum_{i\in\mathbb{K}}x_i$, then 
for any value of $\epsilon > 0$, we have
$\Pr\left(|\mu_{\mathbb{K}}-\mu_{\mathbb{T}}| \geq \sqrt{\frac{n(t+1)}{2t^2(n-t)}\ln{\frac{1}{\epsilon}}}\right) \leq \epsilon.$
\end{lemma}

\noindent{\bf Proof of Theorem~\ref{thm1}}: 
\begin{proof}
We define a random variable $Y_i$ as follows:
  $Y_i= 1$, if $a_i\oplus b_i=x_i \wedge {y_i}$; 0 otherwise.
 Now, we choose a random subset $\Gamma_{CHSH}\subset \{1,\cdots,n\}$ of size $\gamma n$ for any $\gamma>0$ and define $Y=\frac{1}{\gamma n}\sum_{i\in \Gamma_{CHSH}} Y_i$. Here, $Y$ is called observed average value. Let the expected value of $Y$ for that subset be $\mathbb{E}(Y)=\frac{1}{8}(\sin{\theta}(\sin\psi_1+\sin\psi_2)+\cos\psi_1-\cos\psi_2)+\frac{1}{2}$.  Then applying Chernoff bound (Lemma $1$) we get $\Pr\left[|Y-\mathbb{E}(Y)| \geq \delta \right] \leq \exp(-2 \delta^2 \gamma n).$ 

Let $\epsilon_{CHSH}$ be a negligibly small value. Equating $\exp(-2\delta^2 \gamma n)$ with $\epsilon_{CHSH}$ we can find the value of $\delta=\sqrt{\frac{1}{2\gamma n} \ln{\frac{1}{\epsilon_{CHSH}}}}$.
 
 Again, we consider the remaining subset $\Gamma_{QPQ}\subset \{1,\cdots,n\}$ of size $(1-\gamma) n$ and define $Y'=\frac{1}{(1-\gamma)n}\sum_{i\in \Gamma_{QPQ}} Y_i$. Now, from Lemma 3, it can be shown that $\Pr(|Y-Y'|\geq \nu) \leq \exp\left(\frac{-2\gamma^2\nu^2(n-\gamma n)n^3}{(\gamma n+1)n^2}\right).$ 

Let $\epsilon_{QPQ}$ be a negligibly small value. Then, equating the R.H.S with $\epsilon_{QPQ}$, we get $\nu$. \end{proof}

\noindent{\bf Proof of Theorem~\ref{thm2}}: 
\begin{proof}
 In asymptotic limit, i.e., when $n\rightarrow \infty$, the expressions for $\delta$ and $\nu$ tend to $0$. This implies that in asymptotic case, $Y=Y'=\mathbb{E(Y)}$. Thus by calculating the value of $Y$ for the subset $\Gamma_{CHSH}$,  Bob can certify that  entangled states for the subset $\Gamma_{QPQ}$ are of the desired type and hence can be exploited securely for QPQ protocol. Alice can not extract more information about the raw key bits than suggested by the protocol.
 \end{proof}

\end{document}